\documentclass[11pt]{article} 

\usepackage[utf8]{inputenc} 

\usepackage{geometry} 
\usepackage{graphicx} 
\usepackage{array, booktabs, makecell}

\usepackage{booktabs} 
\usepackage{array} 
\usepackage{paralist} 
\usepackage{verbatim} 
\usepackage{subfig} 
\usepackage{gensymb}
\usepackage{mathtools, natbib}
\usepackage{setspace}
\usepackage{caption}
\usepackage{mathtools, natbib}
\usepackage{lscape}
\usepackage{amssymb}
\usepackage{booktabs}
\usepackage{placeins}
\usepackage{xcolor}
\usepackage{chngcntr}

\usepackage[%
    colorlinks=true,
    pdfborder={0 0 0},
    linkcolor=blue
]{hyperref}

\usepackage{fancyhdr} 
\usepackage{setspace}

\usepackage{sectsty}
\allsectionsfont{\sffamily\mdseries\upshape} 

\usepackage[nottoc,notlof,notlot]{tocbibind} 
\usepackage[titles,subfigure]{tocloft} 

\usepackage{lineno}

\usepackage{lscape}

\title{Effects of extending residencies on the supply and quality of family medicine practitioners; difference-in-differences evidence from the implementation of mandatory family medicine residencies in Canada.}

\author{Stephenson Strobel}

\begin{document}
\maketitle

\begin{abstract}

\noindent I examine the impacts of extending residency training programs on the supply and quality of physicians practicing primary care. I leverage mandated extended residency lengths for primary care practitioners that were rolled out over 20 years in Canada on a province-by-province basis. I compare these primary care specialties to other specialties that did not change residency length (first difference) before and after the policy implementation (second difference) to assess how physician supply evolved in response. To examine quality outcomes, I use a set of scraped data and repeat this difference-in-differences identification strategy for complaints resulting in censure against physicians in Ontario. \\

\noindent I find declines in the number of primary care providers by 5\% for up to nine years after the policy change. These changes are particularly pronounced in new grads and younger physicians suggesting that the policy change dissuaded these physicians from entering primary care residencies. I find no impacts on quality of physician as measured by public censure of physicians. This suggests that extending primary care training caused declines in physician supply without any concomitant improvement in the quality of these physicians. This has implications for current plans to extend residency training programs.

\vfill

\end{abstract}

\newpage

\section{Introduction}
\doublespacing

There is interest in extending the length of primary care medical residencies to increase the quality of candidates graduating into independent medical practice. In North America, policy makers are considering extending family medicine residencies by an additional year \citep{fowler_preparing_2022, carek_length_2013, douglass_case_2021, woolever_case_2021}. However, there has been little evaluation of their impact on the supply or quality of physicians.\\

Extending residency might have multiple mechanisms that affect physician supply. There is an obvious mechanical which occurs in the first cohort that has to train for an additional year. When the policy occurs this delays graduation but this disappears once physicians complete the new residency and graduate into independent practice. This ``cohort" effect should have impacts that occur with the policy play out over the  approximate change in length of the residency.\\

However, extending residency programs could have other impacts through altering residency choice. An additional year of training makes other, more lucrative training programs look relatively appealing. This could induce physicians to substitute away from primary care training towards other programs. This ``substitution" effect could potentially have longer-run effects but they would also occur with the onset of the policy.\\

Finally, longer residency lengths could also change the composition of physicians entering primary care which could also have longer run supply effects. If older physicians with shorter work-lives are more likely to choose primary care programs this would mean fewer years of work due to earlier retirement. In contrast with cohort and substitution effects which produce effects coincident with the policy, these composition effects could produce a more pernicious long run supply shock. These would occur at the end of a physicians work-life and would not necessarily be observable in data. \\

Supply effects aside, the stated goal of these primary care residency extensions is to improve quality. If these interventions do cause declines in physician supply, this might be acceptable if there is a trade-off in improved physician quality. However, there is also little evidence on quality changes resulting from program changes. \\

My contribution is two-fold. I estimate the effects of residency training extensions on the supply of physicians. I examine a natural experiment where, over a period of 20 years, Canadian provincial governments mandated two year family medicine residencies in lieu of one year internships for primary care providers. Prior to 1994 medical school graduates had the option to complete a one year rotating internship to become a general practitioner or complete a two year family medicine residency. After 1994 all graduates had to complete a family medicine residency in order to practice in primary care. Using a difference-in-differences (DiD) identification strategy I compare the supply of family medicine practitioners relative to graduates of other specialty programs, like internal medicine, who did not have changes made to their residency lengths (first difference). I use the province-by-province roll out to assess pre-post differences across these groups (second difference).  \\

I then evaluate impacts on quality of residency graduates. Using the same DiD framework comparing specialists to family medicine practitioners I restrict my focus to the province of Ontario. I use a scraped dataset of complaints that result in censures from the College of Physicians and Surgeons of Ontario (CPSO). These censures include terms and conditions imposed on the physician, suspensions and revocation of licence by the CPSO. I can evaluate effects of the change in residency program length on physician quality for over 20 years after the intervention. \\

I find a 5.8\% overall decline in overall numbers of family medicine practitioners after family medicine residencies become mandatory. This occurs over nine years suggesting that a cohort effect cannot fully explain these changes. While these declines in supply disappear by 10 years this is largely in cohorts of physicians who are older and who graduate later. There are persistent declines in cohorts of physicians who have just graduated and in younger physicians suggestive that the policy resulted in primary care physicians that have shorter work-life spans. I find no impact on the quality of physicians as measured by censure by the CPSO. Taken together these results suggest that increasing the length of residencies decreases the supply of independent physicians without any impact on quality of the graduates as measured by complaints resulting in censure.  \\

\subsection{Previous Literature}

I contribute to two major literatures. The first is a medical literature on estimating the supply of physicians. Given long training times and expensive private and public investment \citep{government_of_canada_overview_2009}, most governments and medical organizations project physician supply decades in advance for planning purposes \citep{bureau_of_health_professions_physician_2006, dall_complexities_2021}. A significant amount of policy and academic discourse also centres around how to retain and improve physician supply through incentives \citep{ahmed_looming_2020, golden_managing_2012, chan_perceived_2002, marchildon_physician_2023}. These policies usually involve increasing residency and medical school spots and incentives to retain independent physicians.  \\

Despite the importance placed on securing physician supply, there has been little evidence on how residency length impacts it. Most residents and residency program directors prefer current residency lengths but would consider extensions depending on the scope and content of additional training \citep{duane_length_2002,  smits_residency_2006,  gopal_internal_2007, sabey_views_2015, eiff_comparison_2019}. There is a large discourse consisting of expert opinion predicting the effects of residency extensions \citep{raiche_should_2009, carek_length_2013, hopson_program_2016, douglass_case_2021, woolever_case_2021, glauser_longer_2022}.\\

There is a similar lack of evidence on quality impacts of longer medical residencies. What does exist suggests no association between residency length and baseline applicant quality \citep{eiff_comparison_2019} and no association between residency length and test scores for national certification exams \citep{waller_impact_2017}. Outside of medicine, there is some evidence that reducing high school length leads to poorer post-secondary outcomes \citep{krashinsky_how_2009, meyer_how_2016} and lower wages suggesting poorer productivity related to reduced training \citep{krashinsky_how_2014}. \\

This paper is the first to demonstrate evidence that extending medical residency programs causes decreases in the supply of physicians. These supply effects play out over 10 years after the policy changes. Moreover, based on complaints resulting in censure, there is no improvement in outcomes with longer residencies. \\

This substitution away from primary care residency training links these results to a more general economic literature concerned with intertemporal supply of labour. This literature is based on early models of consumption smoothing \citep{fisher_income_1937}. It has also been extended to examine labour supply over the lifetime which depends on how much agents value leisure, home production and how much their wage changes \citep{mincer_labor_1962, altonji_intertemporal_1986, mulligan_substitution_1998}. Most important to training program selection is that human capital formation is predicted to fall with age because there is less payoff to further investment \citep{ghez_allocation_1975}.\\

From this perspective, additional training time in primary care will make it look less attractive relative to other specialties whose training lengths remain constant. Physicians who switch to take on longer training will be concentrated in younger graduates because they stand the most to gain in earnings over their work-lives. The closest literature that finds similar phenomenon suggests earnings potential can impact choice of university major \citep{arcidiacono_modeling_2012} and can more modestly influence university choice \citep{delavande_university_2019}.\\

However, very little evidence exists on how changes to medical training programs influence specialty choice. Given public interest in physician supply, it is important to understand how changes in policy directly effect supply and how they may change composition in who becomes a primary care physician. If older doctors are entering primary care in lieu of younger physicians this means less years of labour supplied due to shorter work-life spans. I find that evidence that this is the case.\\

\section{Methods}

\subsection{Policy Context}

Canadian medical students apply to residency programs through a centralized matching service which allows them to rank their location and specialty preferences \citep{lim_matchbook_2020}. These include primary care specialties as well surgical, diagnostic, and hospital based specialties. Physician graduates would apply to all types of programs across Canada depending on their location and specialty preferences. Prior to 1997, subspecialized programs with limited positions like neurosurgery would participate in specialized matches outside of the main match. Prior to 1994, medical schools in Quebec did not participate in the centralized match \citep{banner_sandra_pgy-1_1993, banner_sandra_pgy-1_1995, banner_sandra_pgy-1_1997}.  \\

Provincial governments fund and create residency positions for medical students. These are both based on historical and projected future needs. There are other context specific reasons to fund residency spots such as emphasis on French language physicians in Quebec and rural physicians in Northern Canada \citep{dauphinee_what_1997, wilson_what_2017}. As a general rule the number of residency positions exceeds the number of total medical graduates although program or location specific restrictions exist. As an example, during the 1995 match 101 positions across Canada went unfilled with 21 being in family medicine \citep{banner_sandra_pgy-1_1995}. \\

Prior to 1994, medical students had two residency program routes to become independent primary care practitioners.  The first was completing a rotating general internship of one year and then entering independent practice.  Alternately, a medical student could complete a family medicine residency lasting two years. Both routes allowed physicians to practice independently with the former route creating a general practitioner (GP) and the latter route creating a family medicine specialist (FMD). \\

It is worth understanding the flexibility inherent in the rotating internship. Prior to 1993, medical students could enter directly into a major speciality like internal medicine or could spend a year in a rotating internship with a similar first year schedule to all other residencies. This deferred choice for a year and doctors who had completed the rotating residency could then re-enter the match and apply for second year specialty positions. This provided a year of experience for the physician to learn what specialties might be suitable for them. In addition to re-entering the match, physicians had the option of entering primary care practice without further training. Although it continues to be possible to re-enter the match for another program after completing a specialty, it became much less common after the rotating internship was ended. From the period of 1985 to 1995, there was a 40\% reduction in the number of re-entry applicants to residency \citep{dauphinee_what_1997}.\\ 

 However, most provincial governments and the college of family physicians considered a one year period of training after medical school inadequate. Ending the rotating internship occurred on a provincial basis (Table \ref{table_interventions}) and by 1994, the rotating internship had concluded country-wide \citep{levitt_demise_1991, banner_sandra_pgy-1_1995, chan_perceived_2002}.  These changes were grandfathered in for in-practice primary care physicians so that general practitioners who graduated before the policy change did not necessarily have to take additional training and become family physicians. \\

\subsection{Data}

\subsection{The Supply, Distribution and Migration of Physicians in Canada }

I examine the impact of physician supply using provincial level data collected by the Canadian Institute of Health Information. This is recorded annually for each specialty type within a province. I exclude data prior to 1970 due to documentation changes and data after 2003 because of large increases in residency positions \citep{turriff_carms_2020}. \\

This data contains information on the number of physicians and the physicians per capita in each specialty. I am interested in the number of physicians per capita that practice family medicine after the implementation of the policy. I am able to distinguish between GPs who have completed the rotating general internship and FMDs who completed a family medicine residency. I consider the sum of these two categories as the number of primary care practitioners. These data also contain information on the number of physicians by age cohort, and graduation cohort which I exploit in extensions. \\

\subsection{College of Physicians and Surgeons of Ontario Complaints}

To examine the impact on complaints, I use data from the College of Physicians and Surgeons of Ontario. The CPSO is the regulatory body that oversees the professional behaviour of physicians within the province. The CPSO makes data on all members public including any major actions against a physician that result in censure. These data were scraped directly from the CPSO website in 2015 and they contain all individuals who have ever been registered from 1916 to 2015 including individuals who are no longer active.\\

These data include the specialty of the physician and any censure against the physician. These censures, which are the outcome of interest, are whether a license has had terms imposed on it, whether a license has been suspended, or whether a license has been revoked. These do not include complaints which were dismissed as these are not public information. In addition to these concrete measures of a physicians quality, I also examine whether the physician has resigned membership from the CPSO. This is occasionally used as a way to settle a complaint without recording a negative outcome; however, there are also benign reasons for resigning membership such as leaving the province to practice in other jurisdictions. \\

All outcomes except this latter one are relatively rare in the sample of individuals around the policy change (Table \ref{cpso_summary}). Among the 30,823 physicians, only 0.1\% have had a license revoked, 1.9\% have had terms and conditions applied and 0.3\% have been suspended. Family physicians are less likely to have had terms and conditions applied to their license but more likely to have resigned from membership.\\

\subsection{Econometric Strategy}

I use a regression adjusted difference-in-differences (DiD) estimator to assess effects of implementation of residency changes on physician supply \citep{santanna_doubly_2020, callaway_difference--differences_2021}. This approach provides well defined treatment effect parameters as opposed to traditional two-way fixed effects DiDs. I compare the primary care specialties affected to all other physician specialties where training length did not change (first difference). I make this comparison for 10 years pre-post based on the province and the year the policy change occurred (second differences). I use a sample of province-specialties such that 

\begin{equation}
{(Y_{i,j,1}, Y_{i,j,2}...., Y_{i,j,\tau}, D_{i,j,1}, D_{i,j,2}..., D_{i,j,\tau})}^n_{i=1}
\end{equation}

\noindent where $Y$ is the per-capita supply of physicians by specialty $i$ in province $j$ in year $t$. A specialty-province is treated if $D_{i,j,t}=1$ in period $t$. $G_{i,j,g}$ is equal to one when the province-specialty is treated at time $g$ and is otherwise zero.  \\

\noindent Using the never treated province-specialty groups as controls (ie. $C=1$), the average treatment effect on the treated is estimated by:

\begin{equation}
ATT^{nev}_{unc}(g,t)=E[Y_t-Y_{g-1}|G_g=1]-E[Y_t-Y_{g-1}|C=1]
\end{equation}

\noindent This average treatment effect is for treated province-specialties (ie. primary care specialties) at time period $g$ in year $t$. These individual estimates for a time $(g,t)$ are aggregated into a total overall effect for two, four and ten years pre-post to estimate shorter and longer run effects of the policy. I bootstrap standard errors because treatment occurs at the provincial level and there is a few clusters issue \citep{cameron_practitioners_2015, roodman_fast_2019}.\\

DiD models assume that no other interventions occur simultaneously and that in the absence of an effect, outcomes would have trended similarly. The latter assumption can be tested by examining parallel trends in event analyses. Using the above ATT, I can evaluate the impacts for specific points at specific time points around adoption $g$.  All estimates are relative to the year immediately prior to implementation (ie. $g=-1$) and are plotted with 95\% confidence intervals. I limit the window to ten years before and after policy change.  DiDs are implemented via Rios-Avila et al. \citep{rios-avila_csdid_2022}.\\

These events are important for two reasons. The first is a test on pre-trends which provides evidence of causality. The second is that they suggest mechanisms. If the effect of the policy change is purely through a mechanical impact on the cohort of entering family physicians these should evolve over the new period of residency length. If effects persist beyond this period, it suggests that supply is impacted because of a decline in individuals entering into family medicine residencies. This is consistent with a substitution effect where physicians choose other specialties.\\

I use a similar econometric strategy to assess whether physicians receive more complaints as a result of the increase in residency length in Ontario. This collapses to a canonical DiD with one point of treatment in 1994. This is specified as 

\begin{equation}
    c_{ijt}= \beta_i(D_{ijt}*Post_{ijt})+G_{ij}+T_{t}+\epsilon_{ij}
\end{equation}

\noindent where a dummy variable, $c$, is equal to one if a complaint has resulted in a public censure for a physician $i$ in specialty $j$, who graduated in year $t$. $D$ is equal to one for specialties that are treated which are total family practitioners, FMDs and GPs. $Post$ is equal to one for these specialties in Ontario after implementation and zero for other specialties. I am interested in the effects of $\beta_i$; this is the overall difference-in-differences effect on physician quality. $G$ is a fixed effect for the specialty group. $T$ is a fixed effect for the year of graduation.  $\epsilon$ is a bootstrapped error term.\\

Event regressions are specified as

\begin{equation}
    y_{ijt}=\alpha_i \sum^{10}_{-10} (D_{ijt}*Post_{ijt})+G_{ij}+T_{t} +\epsilon_{ij}
\end{equation}

\noindent $\alpha_i$ estimates effects for each individual dummy variable. These dummies take a value of one for family medicine related specialty groups in the years after 1994 when there was implementation of mandatory family physician training in Ontario. They are zero otherwise. All estimates are relative to the year immediately prior to implementation (ie. $t=-1$) and are plotted with 95\% confidence intervals. I limit the window to ten years before and after policy change.\\

\subsection{Extensions: Age of physician and physician year of graduation}

The overall impact that I estimate occurs across all physicians regardless of age or graduation year. Changes to this overall number of physicians could be from new graduates, retirees, or physicians who migrate. The CIHI data has information on the age and age cohorts of family physicians in practice in a province in a given year which is suggestive of who is affected. As the demographics of physician graduates skew younger \citep{lakoff_analysis_2020}, if residency classes are affected this should increase the average age of primary care providers. To demonstrate that this overall supply effect is likely coming through changes to graduating residents I repeat the above DiD estimation strategy using average age as an outcome. Similarly, I examine the supply of ``young" primary care physicians as defined as those under 40, and ``old" family physicians as defined as those 60 and older. I am also able to perform a similar exercise by year of graduation. If this effect comes through changes in residency, cohorts who are closer to medical school graduation should be more affected.\\

\subsection{Extensions: Consistency of effects across the four interventions.}

Residency choice is a mixture of specialty preferences and location preferences. In provinces that adopt the policy earlier, medical school graduates could leave the province to complete a shorter residency. To assess whether main effects are consistent I examine the impact of the policy changes by year of adoption. By the 1994 cohort, there are no other provinces where graduates could move in order to avoid longer residencies.\\

\subsection{Extensions: Substituting towards other specialties}

If declines in physician supply are purely through a cohort effect, then supply will be constrained for a limited period but eventually rebound. However, if the effect is through a substitution effect, supply may be constrained for a much longer time with more pernicious effects. Two results might suggest that medical students are substituting towards other programs. First, if changes in primary care physician volumes occur more than two years after implementation (ie. the length of that residency), it suggests that delayed cohort effects are not the only mechanism. Second, specialties that attract similar candidates as family medicine should see increases in their graduating residents. \\

There is little evidence on the alternate preferences for individuals who are admitted to family medicine programs. However there is guidance from medical student groups which suggest specialties that are similar to family medicine in emphasis. These are internal medicine, pediatrics, neurology, psychiatry, and physical medicine \citep{lim_matchbook_2020}. I repeat the above DiD and event exercise with these as the treatment groups.\\

\subsection{Extensions: Altering the control group}

Along similar lines as the previous extension, I examine effects when changing the control group of specialties. I change this to a group of specialties that are thought to have relatively low substitutability with family medicine. These are specialties that have a diagnostic and surgical focuses.\\

\section{Results}

\subsection{Effects on physician supply}

\begin{figure}[h!]
\centering
\includegraphics[width=150mm]{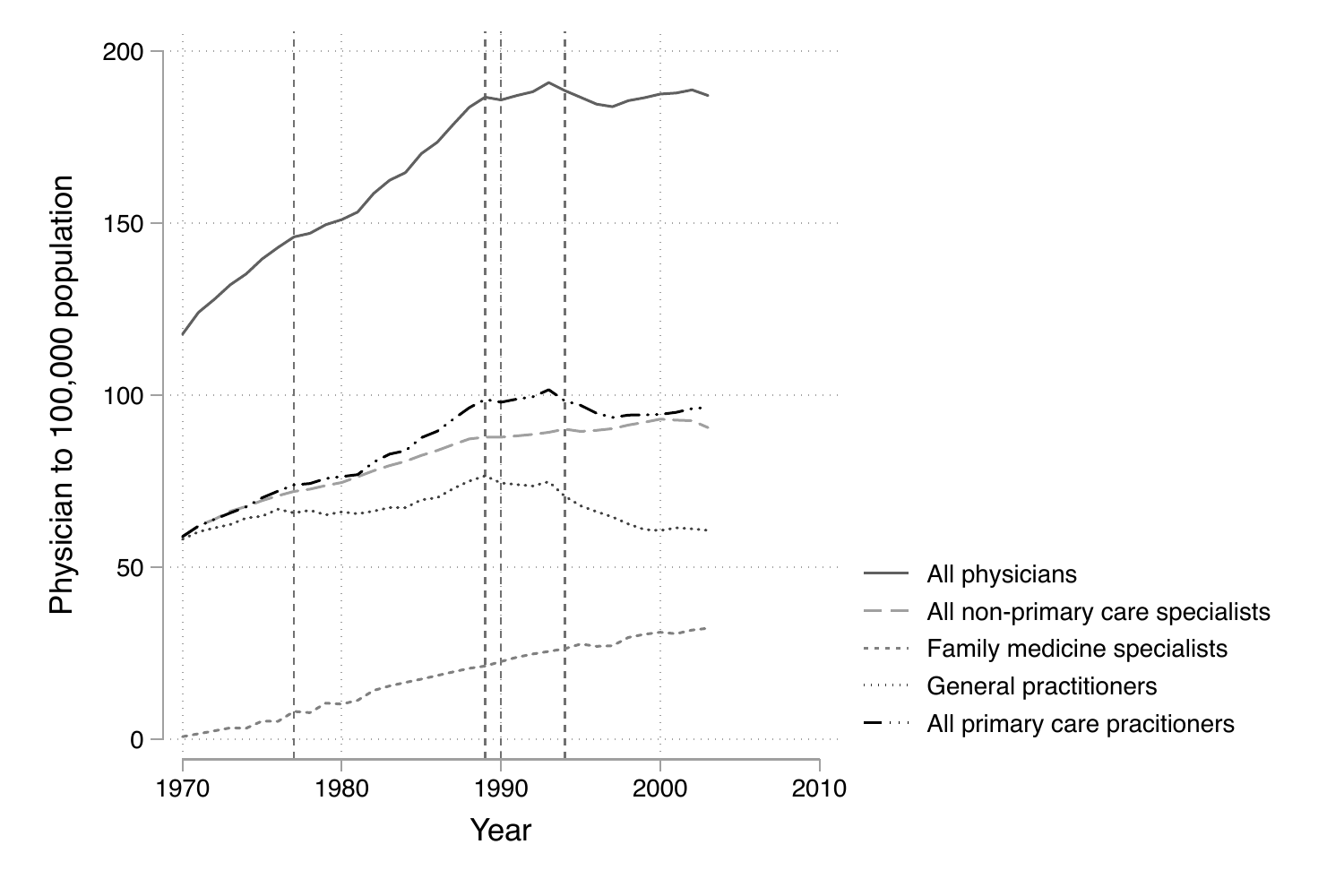}
\caption{Time series of selected physician categories. Lines indicate dates of provincial changes to residency program length.}
\label{time_series}
\end{figure}

Figure \ref{time_series} demonstrates the time-series of selected types of physicians over the period of 1970 to 2003. Around 1994 when rotating internships are completely phased out in Canada, the number of GPs begins to decline from 75 per 100,000 population. The number off FMDs increases and in 1994 is approximately 26 per 100,000 population. The overall number of primary care practitioners in 1994 is approximately 100 per 100,000 population.\\ 

Table \ref{table_dids} demonstrates the DiD effects estimated by total primary care practitioners, FMDs, and GPs. There is an decrease in the overall supply of family medicine practitioners but at statistically insignificant levels. This effect is negative at marginally statistically significant levels for shorter time horizons but zero for a 10 year horizon. At the 10\% level the policy causes a 1.56 and 1.89 total primary care provider reduction per 100,000 persons at two and four years respectively. \\

\begin{figure}[h!]
\centering
\includegraphics[width=150mm]{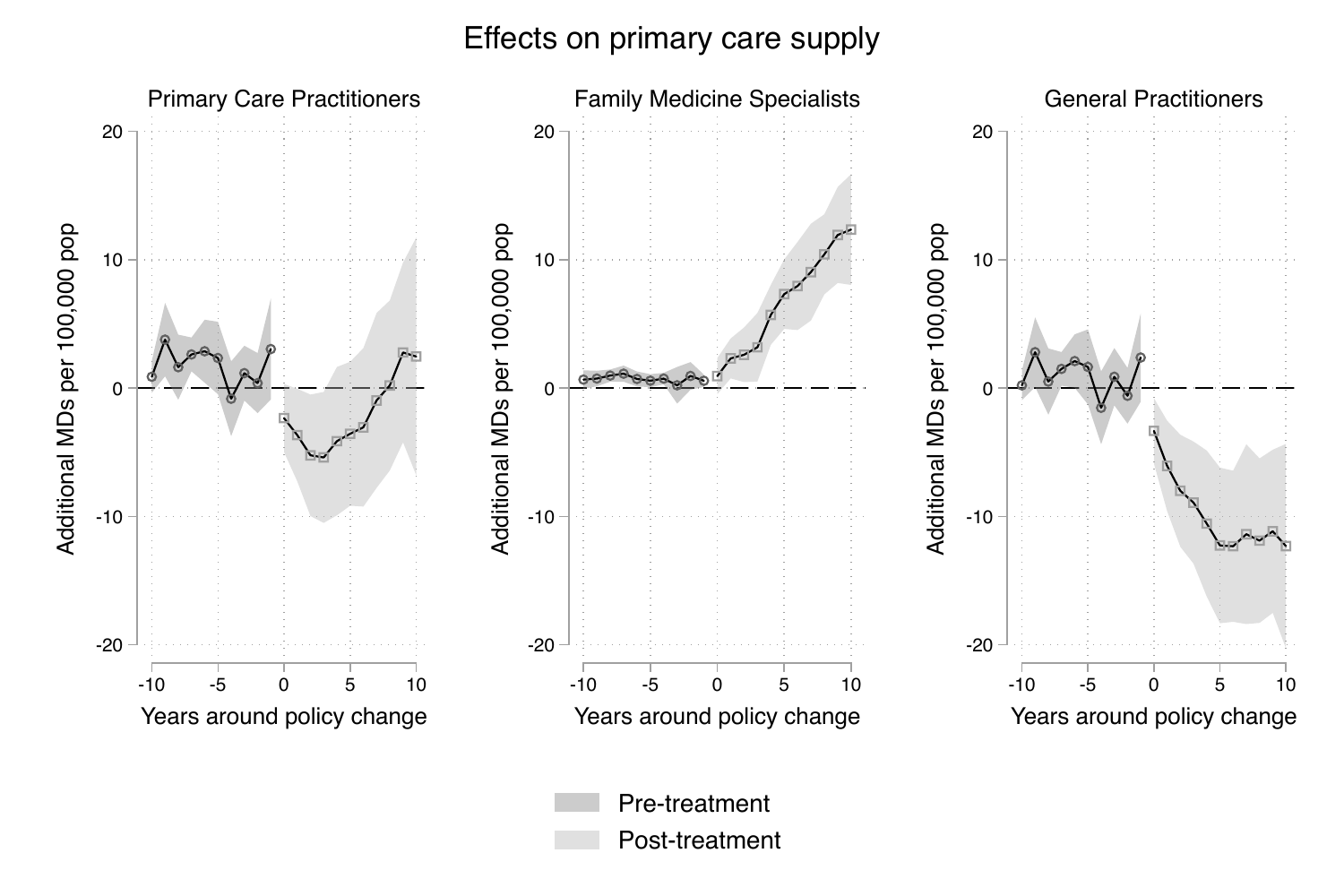}
\caption{Event analysis for family medicine practitioners per 100,000 persons. Point estimates are displayed with bootstrapped 95\% CI.}
\label{supply_physicians100}
\end{figure}

\begin{figure}[h!]
\centering
\includegraphics[width=150mm]{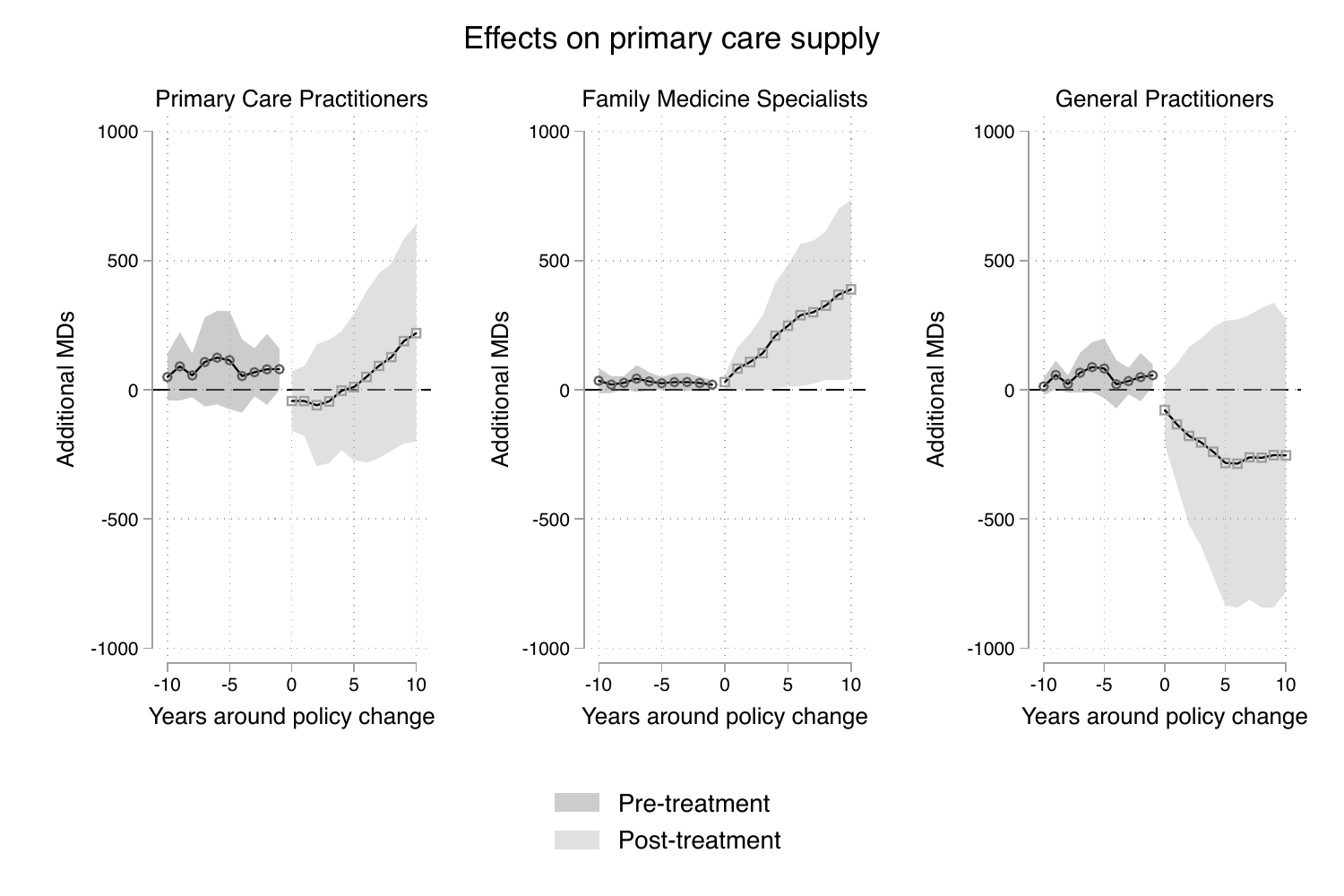}
\caption{Event analysis for total number of family medicine practitioners. Point estimates are displayed with bootstrapped 95\% CI.}
\label{supply_physicians}
\end{figure}

Figure \ref{supply_physicians100} demonstrates the estimates from event analyses by type of physician per 100,000 population. Pre-trends in all graphs are noisy but relatively flat and stable suggesting that the parallel trends assumption holds well. In the period after mandating increased training there are large declines in the number of GPs and large increases in the number of FMDs. However, there is a transient trough in supply about five years after policy implementation when the net effect is negative and statistically significant. That trough demonstrates a reduction in the number of total family practitioners of 5-6 per 100,000 population. As an alternate specification, I demonstrate effects on the physician supply irrespective of population in figure \ref{supply_physicians}. These estimates are much noisier but demonstrate similar point reductions for five  years after implementation. This decline in the overall number of primary care practitioners is an approximately 20 per year difference between the pre-post periods.\\

\subsection{Extensions}

\begin{figure}[h!]
\centering
\includegraphics[width=150mm]{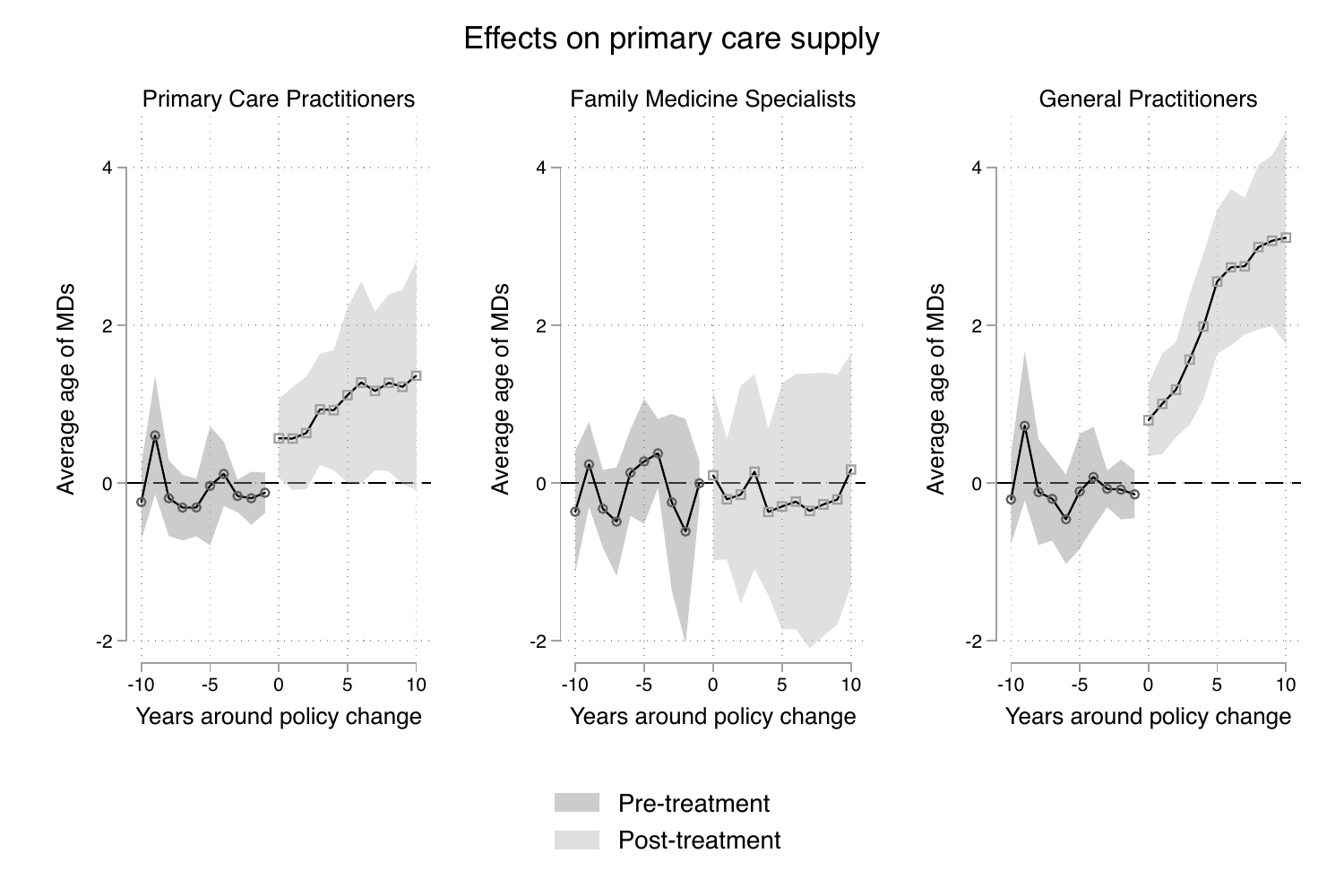}
\caption{Event analysis for the average age of primary care practitioners. Point estimates are displayed with bootstrapped 95\% CI.}
\label{supply_avgage}
\end{figure}

\begin{figure}[h!]
\centering
\includegraphics[width=150mm]{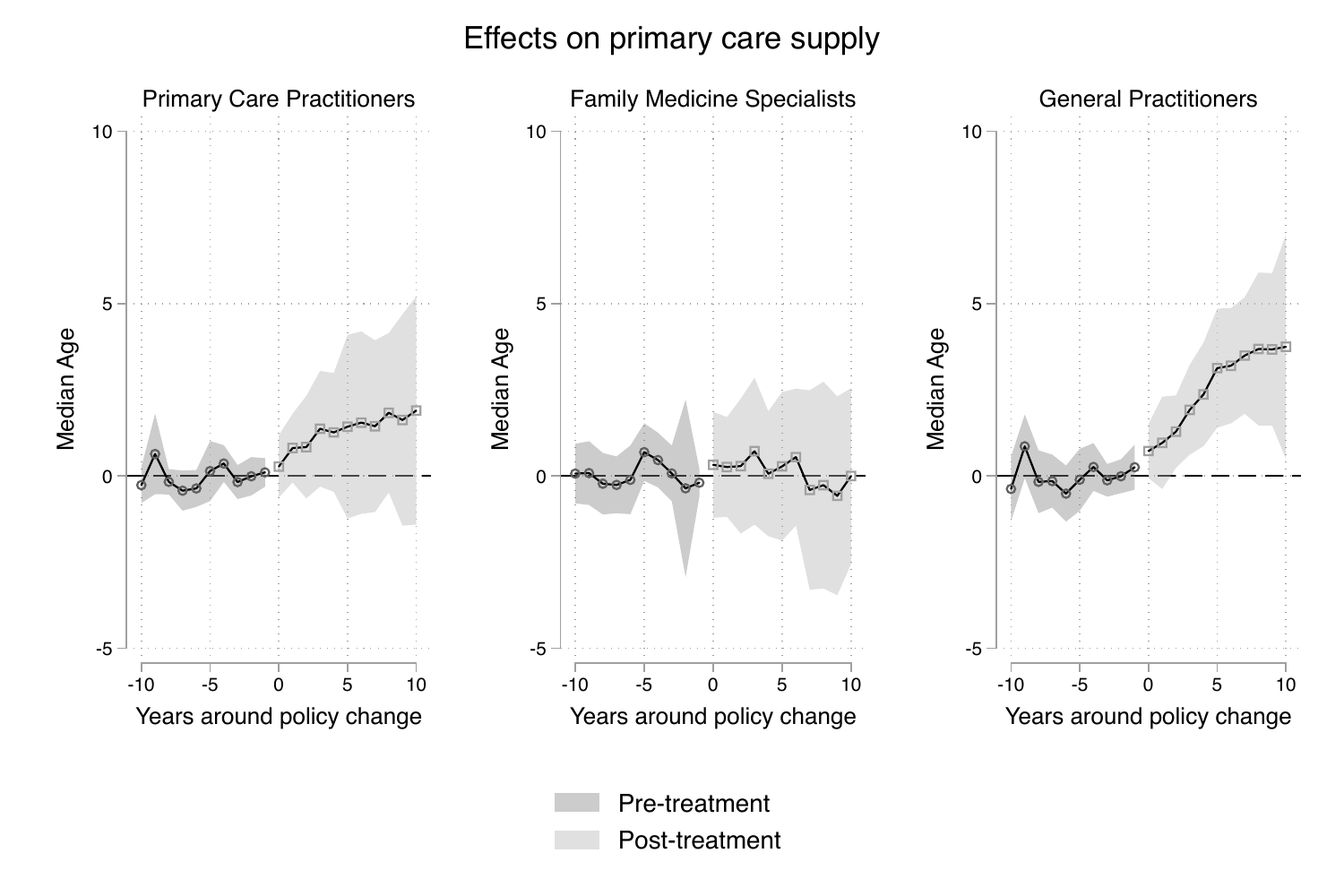}
\caption{Event analysis for the median age of primary care practitioners. Point estimates are displayed with bootstrapped 95\% CI.}
\label{supply_medage}
\end{figure}

 Figure \ref{supply_avgage} demonstrates effects on the age of primary care practitioners. There is an initial step increase in the average age of family practitioners. While there are no increases in average age of family medicine specialists there are increasing average ages of general practitioners. This is consistent with blocked entry into this route to practice primary care. However the overall average age of all primary care practitioners continues to increase. Event studies of the median age show similar effects (Figure \ref{supply_medage}).\\

\begin{figure}[h!]
\centering
\includegraphics[width=150mm]{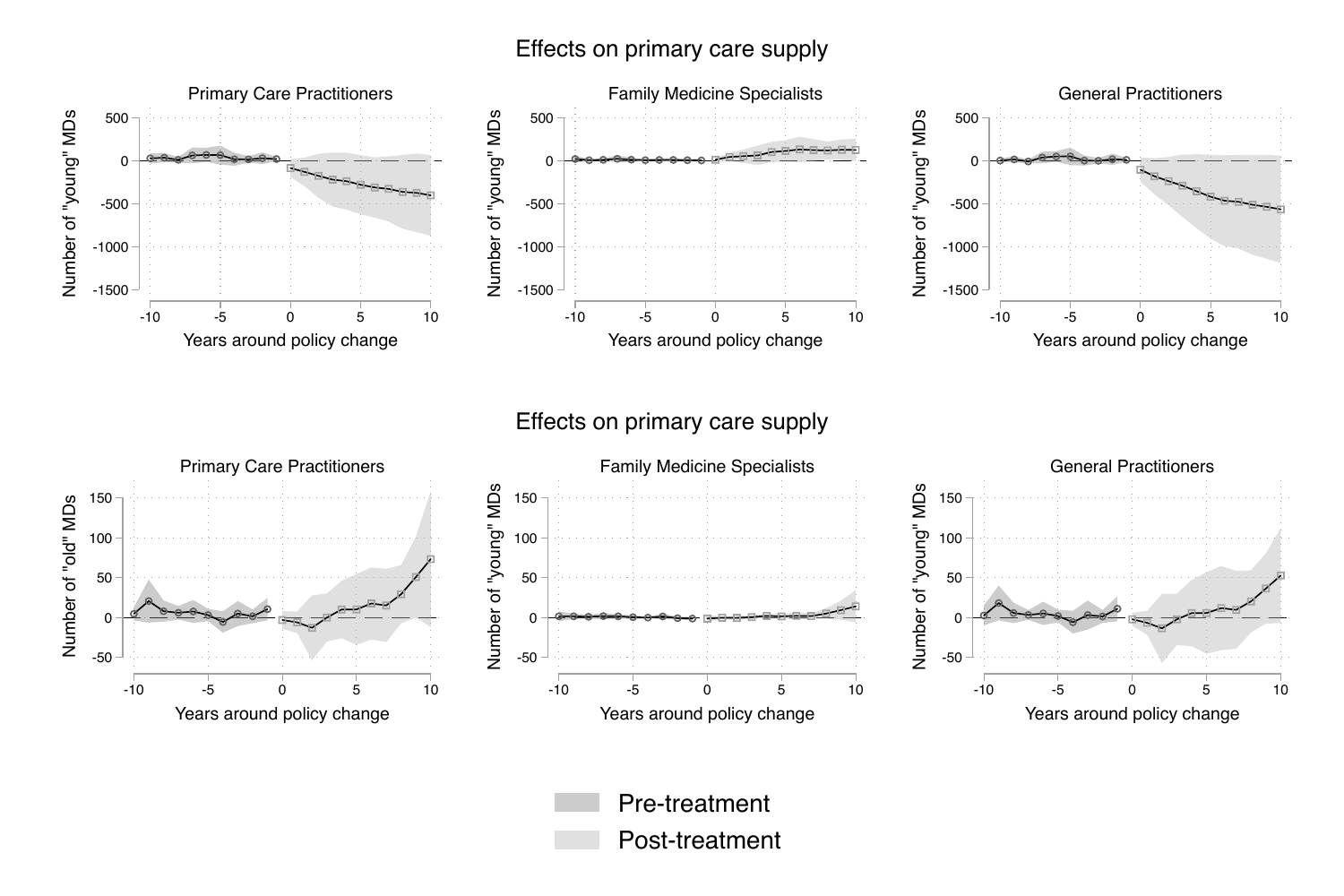}
\caption{Event analysis for ``young" and ``old" primary care practitioners.``Young" MDs are any physician under the age of 40 whereas "old" MDs are any physician over the age of 60. Point estimates are displayed with bootstrapped 95\% CI.}
\label{supply_young_old}
\end{figure}

These results are similar when examining the number of physicians by age cohort (Figure \ref{supply_young_old}). This policy does not affect cohorts of ``old" physicians until the end of the 10 year period of observation when younger general practitioners are aging into the ``old" cohort. However, the policy has a relatively large and immediate effect on cohorts of ``young" physicians. Although noisily estimated, by the end of the 10 year period there are 200 fewer family medicine practitioners per year under the age of 40. The overall 10 year supply of these ``young" MDs is reduced by 120 primary care practitioners with an additional 58.57 family medicine specialists but 188 fewer general practitioners (Table \ref{dids_time}). \\

\begin{figure}[h!]
\centering
\includegraphics[width=150mm]{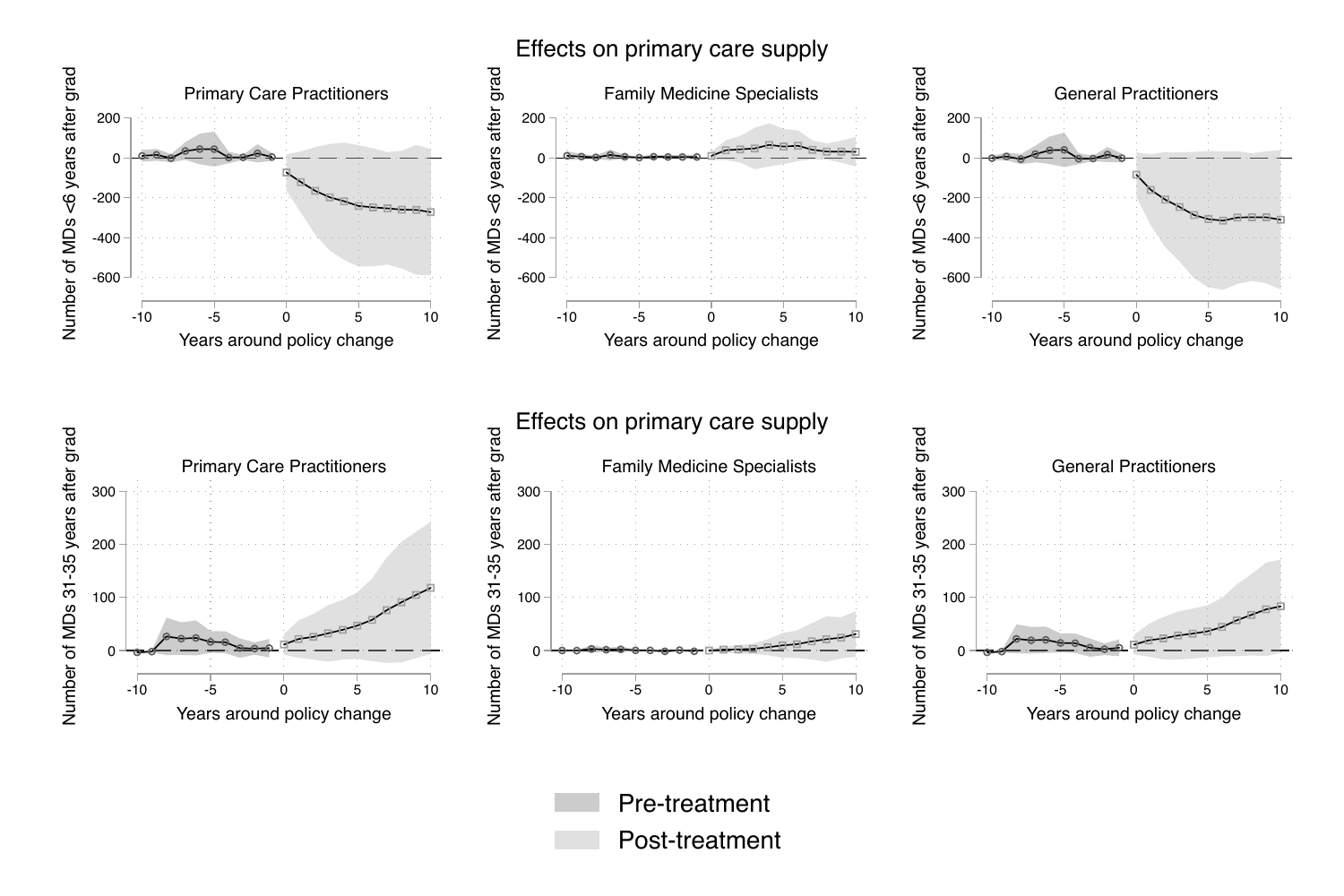}
\caption{Event analysis for example graduation cohorts. Point estimates are displayed with bootstrapped 95\% CI.}
\label{supply_graduation}
\end{figure}

Reinforcing this, declines in physician supply resulting from the policy are through cohorts of more recent medical school graduates (Figure \ref{supply_graduation}). Over 10 years, the number of primary care practitioners who graduated from medical school within six years declines by 103 per year and this effect occurs immediately suggesting that fewer overall recent grads are entering primary care. This occurs through a decline in general practitioners by 131, with an increase of 25.6 in the number of family medicine specialists (Table \ref{dids_time}).\\

\begin{figure}[h!]
\centering
\includegraphics[width=150mm]{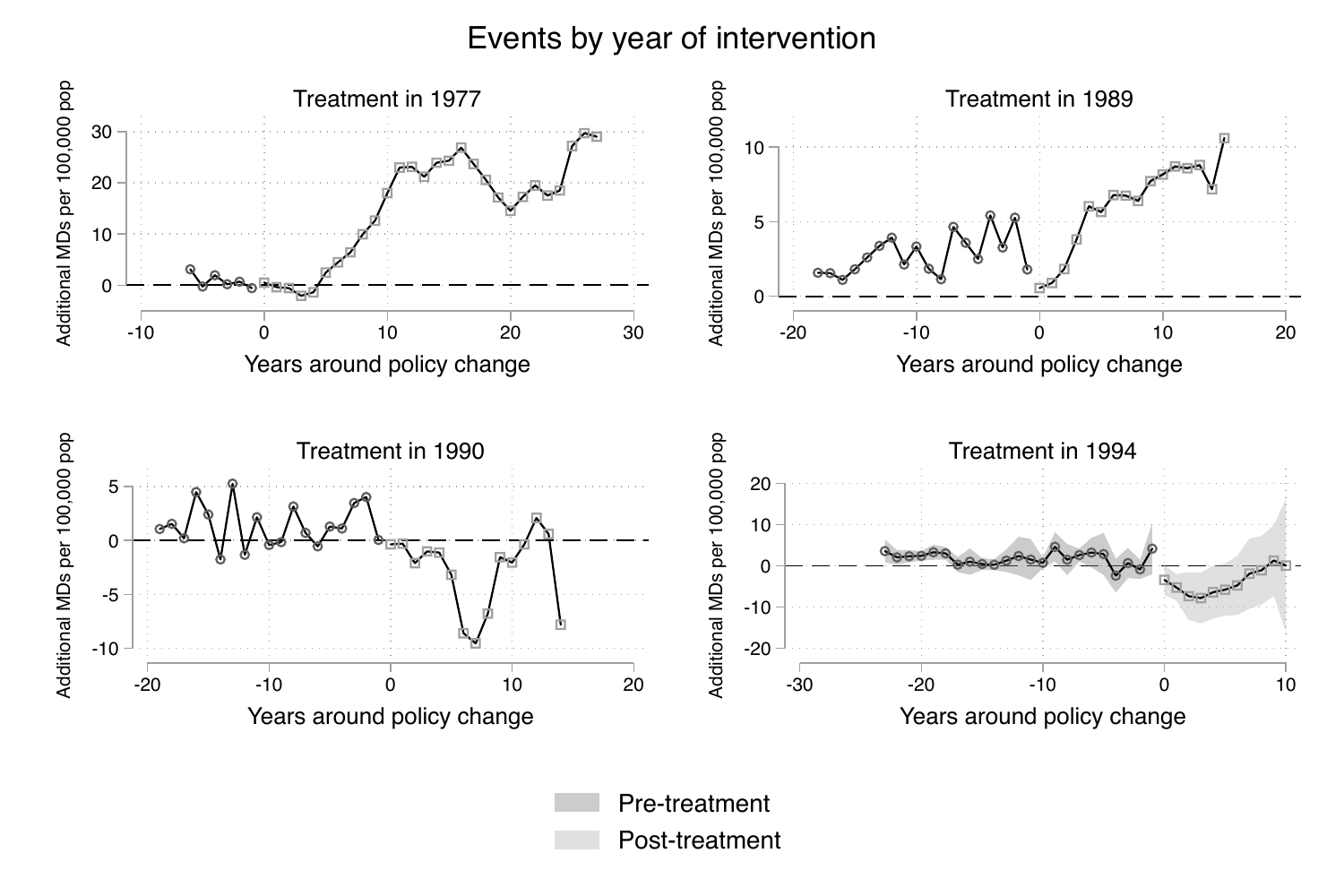}
\caption{Event analysis by year of intervention. Point estimates are displayed with bootstrapped 95\% CI.}
\label{supply_cohort}
\end{figure}

Figure \ref{supply_cohort} demonstrates effects by year of intervention. Short-run effects are relatively consistent across all four intervention times. Pre-trends are relatively stable in cohorts except for the 1989 intervention which occurred in Quebec. \\

\begin{figure}[h!]
\centering
\includegraphics[width=150mm]{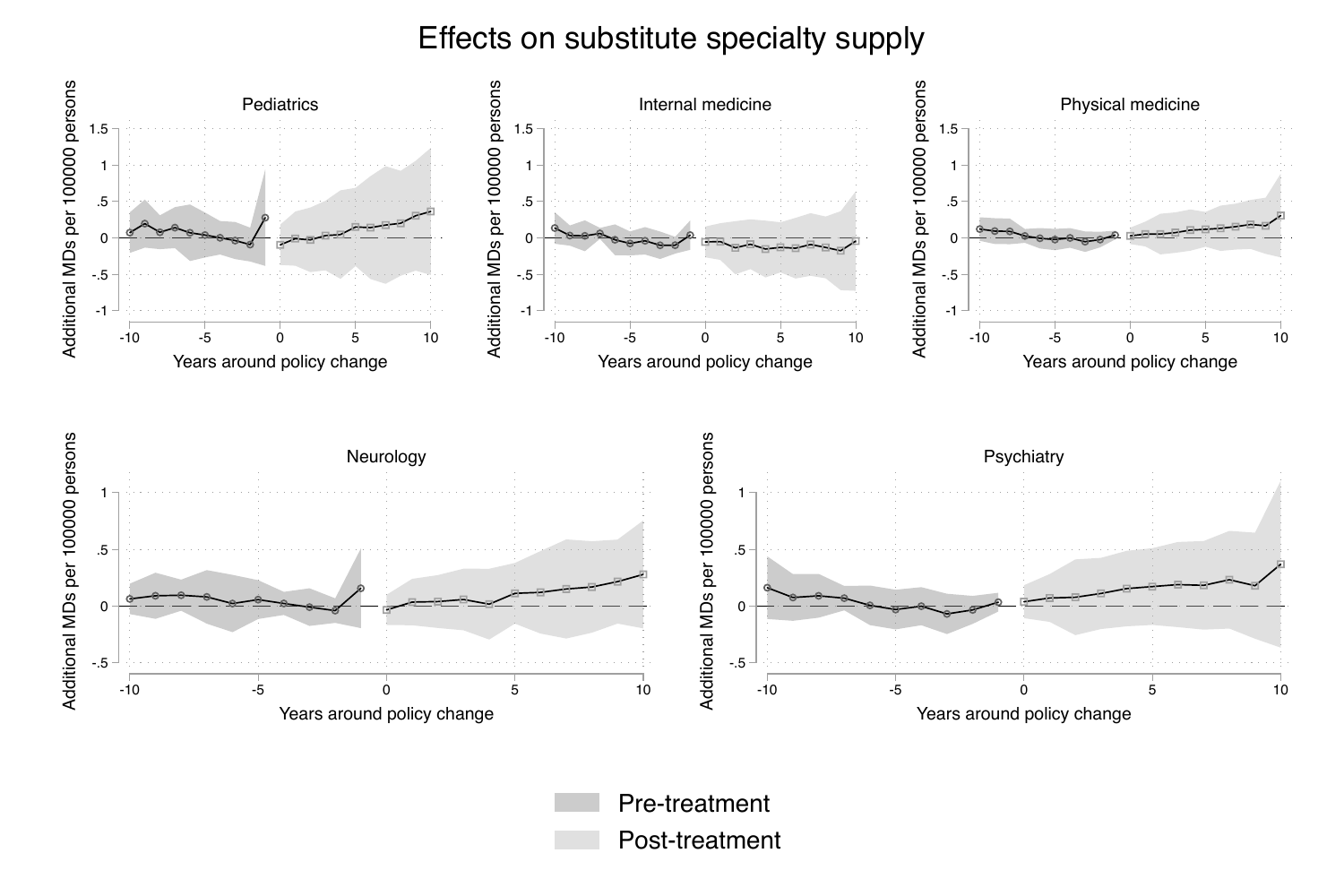}
\caption{Event analysis for substitute specialties to family medicine. Point estimates are displayed with bootstrapped 95\% CI.}
\label{supply_alternative}
\end{figure}

Figure \ref{supply_alternative} demonstrates the effects of the policy on substitute residency programs. There are trends towards increases in these residencies over the 10 years after the policy change.  These effects peak about five years after the policy change which is the usual length of a specialist residency in Canada. Table \ref{table_dids} also demonstrates the DiD estimates for these specialties. \\ 

\begin{figure}[h!]
\centering
\includegraphics[width=150mm]{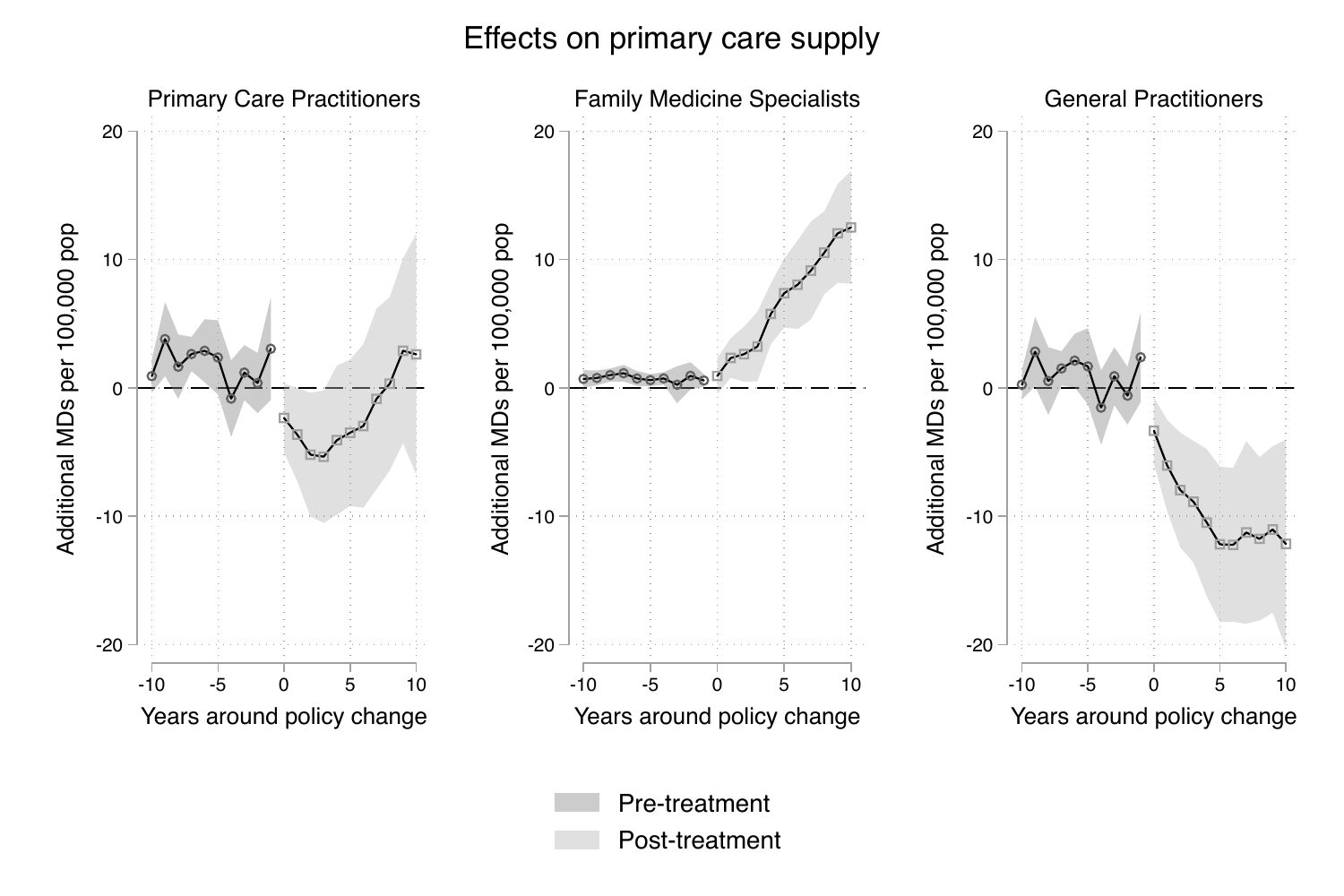}
\caption{Event analysis for total number of family medicine practitioners per 100,000 population relative to an altered control group. This control group consists of surgical and diagnostic residency specialties. Point estimates are displayed with bootstrapped 95\% CI.}
\label{supply_control}
\end{figure}

These effects are stable to alteration of the control group. Figure \ref{supply_control} demonstrates event analyses when the control group consists solely of surgical and diagnostic specialties. Table \ref{table_dids2} demonstrates DiD estimates.\\

\subsection{Effects on physician quality}

\begin{figure}[h!]
\centering
\includegraphics[width=150mm]{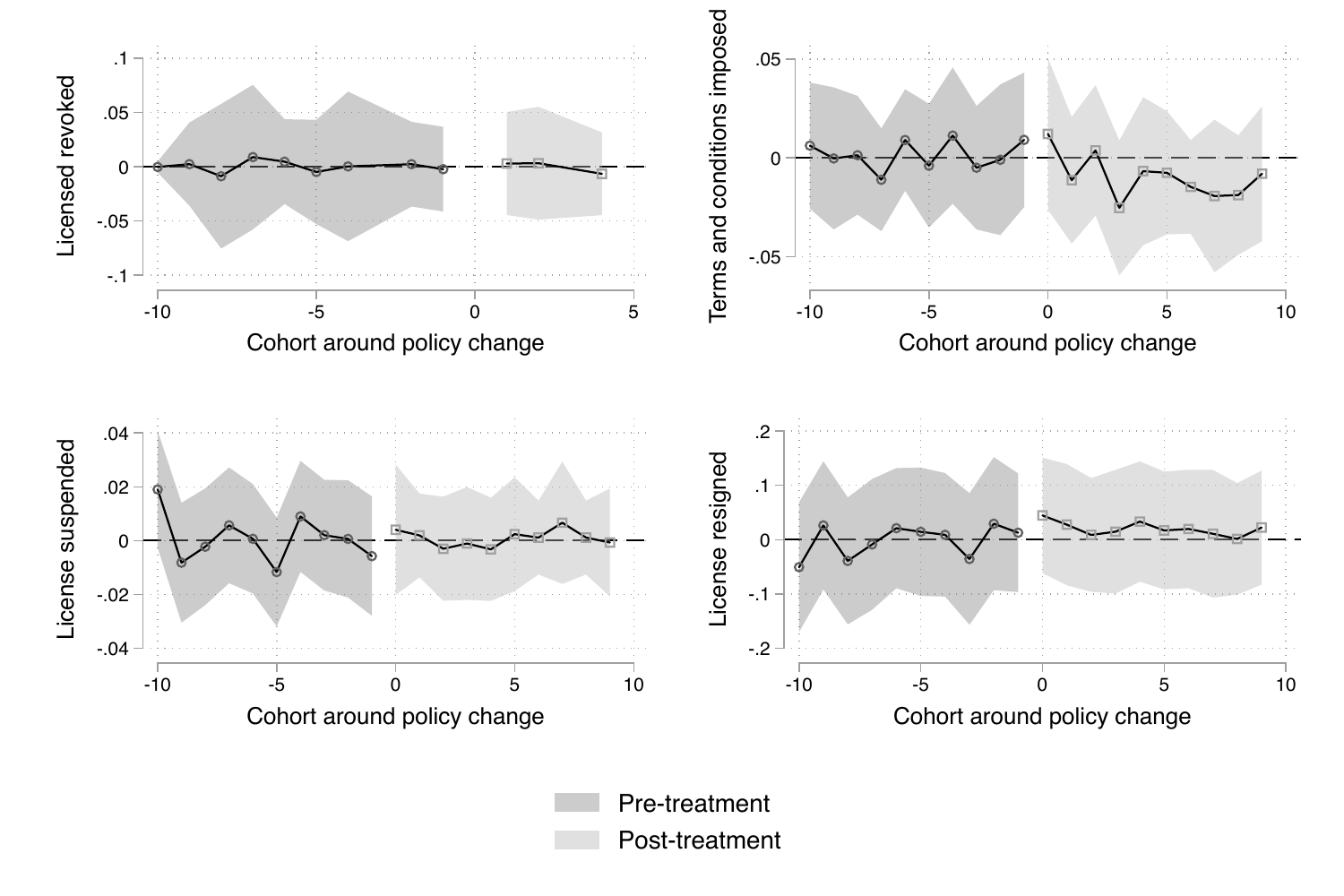}
\caption{Event analysis for quality measures in Ontario for primary care practitioners. Point estimates are displayed with bootstrapped 95\% CI. Note that revocation is such a rare quality outcome that estimates cannot be produced for several years.}
\label{cpso_events}
\end{figure}

Table \ref{cpso_quality} demonstrates the DiD estimates over a two, four and ten year horizon. There are no significant effects save for a marginal increase in the probability of resigning from Ontario membership in the cohorts graduating after extensions of family medicine residencies are implemented. Figure \ref{cpso_events} demonstrates the events of these effects. There are similar zero estimates on probability of negative outcomes for graduating cohorts after the policy change.\\

\section{Discussion}

I leverage the phasing out of the rotating medical internship in Canada to demonstrate the effects of a change in length of residency on physician supply and quality. This effectively increased a primary care residency from one to two years of training. This policy caused a transient decline in the total supply of primary care practitioners. At its trough, this is a reduction by 5.8\% of the overall number of practitioners. Current projections suggest that supply of primary care specialists will need to increase by 7 to 20\% by 2034 to keep up with demand \citep{dall_complexities_2021}. These results suggest that extending family medicine residencies will cause an additional decline in supply of physicians by the lower bound of these estimates.\\

These results do not occur solely because of a delays of the initially affected cohort's entry into independent practice. The measured decline is statistically significant and negative for four years after intervention and point estimates remain negative for nine years. This trough in supply is several years longer than the length of a family medicine residency which is not consistent with a pure cohort effect of lengthening the residency. While I cannot observe direct substitution effects, there is some evidence consistent with substitution towards other specialties occurring in historical CaRMS reports from 1995. Prior to the transition, medical school graduates could expect to match to one of their top three programs 80\% of the time; in 1993, this dropped to 70\% suggesting that there was poorer matching success around the time when the majority of provinces discontinued the rotating internship \citep{banner_sandra_pgy-1_1995}. \\

These policies also create a change in the composition of primary care practitioners with possible long-run supply effects. Younger cohorts and earlier graduates are the physicians most likely to avoid primary care after the increase in program length. This creates a possible decline in long-run supply. Primary care physicians trained after the change are older and so may work fewer years during their career. \\

However, these results should be interpreted in the context of the other changes that occurred with abolishing the rotating internship. This not only extended the training required to become a credentialed primary care practitioner but also meant reduced flexibility in program choice. In many cases, the rotating internship was not an absorbing state and was used as a way for a physician to test which specialty they would apply to in subsequent re-entry matches. It was also often the case that general practitioners would practice several years before going into a non-primary care specialty \citep{dauphinee_what_1997}.\\

It is reassuring that policy effects are relatively consistent across intervention years (Figure \ref{supply_cohort}) especially when flexibility existed to switch to provinces that retained the rotating internship; nevertheless, this provides two points of caution in applying these results. First, extending primary care residencies today would not be contaminated by these dual effects of flexibility and extended length. It is ambiguous as to how this would change results. Second, the implied effects that occur because of changes in the work-life length of primary care physicians depends on how absorbing primary care was for a general practitioner. More evidence on this point is needed.\\

These results also say nothing about access; it is possible that primary care practices were able to absorb patients who could not access new family practitioners. Moreover, residents entering family medicine programs treat patients and often work very long hours \citep{pattani_resident_2014} which could potentially increase access.  Accounts of the policy change in 1994 also suggest that this altered where patients could receive resident based care implying changes in the distribution of care; it reduced trainee availability in hospitals and improved access at community sites \citep{dauphinee_what_1997}.  \\

This access to additional trainees is balanced against how autonomous the resident is in patient care. Residents often perceive a lack of supervision \citep{baldwin_how_2010, mieczkowski_perceptions_2021} which is consistent with the resident having some ability to produce in parallel to independent practitioners. However, in at least one resident setting, 30\% of clinics required faculty to see each patient treated by a resident and 42\% had to review each patient \citep{singman_supervision_2017}.\\

These overall impacts on supply may also reflect contaminated effects. The majority of provinces abolished the one year internship in the 1990s when other major changes were being made in physician training. Provincial policies towards medical training were heavily influenced by reports that suggested that there were too many medical school graduates \citep{marchildon_physician_2023}. Provincial governments cut these positions by up to 10\% in the years after 1991 although this did not lead to any reduction in residency positions \citep{dauphinee_what_1997}. This coincided with federal fiscal retrenchment in 1995 which reduced federal health transfers to provinces. By 1997, this resulted in a physician shortage which triggered provincial governments to provide more funding for training. This timing roughly corresponds with these policy changes and might explain why physician supply might rebound in the event analyses.\\

These declines in supply may be an acceptable trade-off if primary care quality improves because of the longer training. To the extent that censure by a licensing body represents physician quality, I find no improvements in outcomes in Ontario caused by the policy change. More evidence on quality improvement is needed as these results only speak to one possible avenue where quality might be affected. This result though suggests that supply may be impacted without improvements in quality.\\
 
In sum, these results show that extending the length of a family residency program in Canada led to a reduction in the supply of primary care physicians. Policy makers should consider these impacts on supply and quality in the current debate on proposed extensions to family medicine training.\\

\newpage
\bibliographystyle{apalike}
\bibliography{DoctorsNotch.bib}

\newpage
\section{Tables and Figures}
\FloatBarrier

\begin{table}
\begin{center}
\begin{tabular}{l|c|c} \hline
Province                      & Year of policy &  Year of Effect \\  \hline \hline
Alberta        & 1976           & 1977                 \\ 
Quebec                        & 1988           & 1989                 \\
Saskatchewan                  & 1989           & 1990                 \\
British Columbia              & 1993           & 1994                 \\
Ontario                       & 1993           & 1994                 \\
New Brunswick                 & 1993           & 1994                 \\
Prince Edward Island          & 1993           & 1994                 \\
Nova Scotia                   & 1993           & 1994                 \\
Manitoba                      & 1994           & 1994                 \\
Newfoundland and Labrador     & 1994           & 1994    \\            \hline
\end{tabular}
\caption{Year of policy change by province. Note that Manitoba and Newfoundland and Labrador implemented policies abolishing rotating medical residencies in 1994 but these affected the residency class of 1992-1993.}
\label{table_interventions}
\end{center}
\end{table}

\begin{landscape}

\begin{table}[h!] \scriptsize \centering
\def\sym#1{\ifmmode^{#1}\else\(^{#1}\)\fi}
\begin{tabular}{l|c c c | c c c c c}
\hline\hline
            &\multicolumn{3}{c|}{Primary Care Practitioners} &\multicolumn{5}{c}{Substitute Specialties}\\
            \hline
            &\multicolumn{1}{c}{Total}&\multicolumn{1}{c}{FMDs}&\multicolumn{1}{c|}{GPs}  &\multicolumn{1}{c}{Pediatrics}&\multicolumn{1}{c}{Neurology}&\multicolumn{1}{c}{Internal Medicine} &\multicolumn{1}{c}{Psychiatry} &\multicolumn{1}{c}{Physiatry}\\
\hline
DiD Estimates (2 Years)        &      -1.563\sym{+}&       1.468\sym{*}&      -3.124\sym{*}  &         0.0114       &      0.0319       &     -0.0601       &      0.0379       &      0.0303     \\
             &     (-1.85)       &      (4.05)       &     (-3.78)      &      (0.13)       &      (0.70)       &     (-0.97)       &      (0.75)       &      (0.76)        \\
\hline
DiD Estimates (4 Years)    &    -1.891\sym{+}&       1.908\sym{*}&      -3.972\sym{*} &        0.0114       &      0.0277       &     -0.0754       &      0.0429       &      0.0312      \\
            &     (-1.83)       &      (4.70)       &     (-3.78)   &       (0.12)       &      (0.61)       &     (-1.19)       &      (0.81)       &      (0.78)       \\
\hline
DiD Estimates (10 Years)  &    -0.288       &       3.912\sym{*}&      -4.796\sym{*}  &      0.0968       &      0.0814       &     -0.0610       &      0.0995       &      0.0785      \\
           &     (-0.27)       &      (8.19)       &     (-4.45)     &        (0.84)       &      (1.45)       &     (-0.87)       &      (1.41)       &      (1.41)             \\
\hline
\(N\)       &           8690        &         8690   &       8690   & 8350 & 8010 & 7704 &7364 & 7024    \\
Mean        &       84.34       &       16.34       &       67.35  &  5.292       &       3.328       &       4.929       &       5.563       &       4.620      \\
\hline\hline
\multicolumn{9}{l}{\footnotesize \textit{t} statistics in parentheses}\\
\multicolumn{9}{l}{\footnotesize \sym{+} \(p<0.10\), \sym{*} \(p<0.05\)}\\
\end{tabular}
\caption{DiD Estimates by type of physician.}
\label{table_dids}
\end{table}

\begin{table}\scriptsize \centering
\def\sym#1{\ifmmode^{#1}\else\(^{#1}\)\fi}
\begin{tabular}{l |ccc|ccc|ccc}
\hline\hline
            &\multicolumn{3}{c}{2 Year Effects} &\multicolumn{3}{c}{4 Year Effects} &\multicolumn{3}{c}{10 Year Effects}\\ \hline
            &\multicolumn{1}{c}{Total}&\multicolumn{1}{c}{FMDs}&\multicolumn{1}{c}{GPs} &\multicolumn{1}{c}{Total}&\multicolumn{1}{c}{FMDs}&\multicolumn{1}{c}{GPs} &\multicolumn{1}{c}{Total}&\multicolumn{1}{c}{FMDs}&\multicolumn{1}{c}{GPs}  \\
\hline
Average Age         &       0.290\sym{*}&      -0.174       &       0.552\sym{*}&       0.362\sym{*}&      -0.107       &       0.701\sym{*}&       0.495\sym{*}&      -0.126       &       1.123\sym{*}\\
            &      (2.10)       &     (-0.47)       &      (4.57)       &      (2.60)       &     (-0.39)       &      (5.04)       &      (2.57)       &     (-0.40)       &      (7.84)       \\ \hline
Young MDs         &      -64.15       &       28.21\sym{*}&      -94.55\sym{+}&      -80.96       &       37.97\sym{*}&      -122.2\sym{*}&      -120.5\sym{+}&       58.57\sym{*}&      -188.5\sym{*}\\
            &     (-1.34)       &      (2.09)       &     (-1.84)       &     (-1.40)       &      (1.96)       &     (-1.97)       &     (-1.82)       &      (2.22)       &     (-2.16)       \\ 
Old MDs         &      -1.755       &      -0.575       &      -1.775       &       0.186       &       0.219       &      -0.825       &       11.97\sym{+}&       2.210\sym{+}&       8.045       \\
            &     (-0.35)       &     (-0.71)       &     (-0.41)       &      (0.03)       &      (0.39)       &     (-0.14)       &      (1.84)       &      (1.80)       &      (1.29)       \\ \hline
Graduation - Fewer than 6 years         &      -65.86\sym{+}&       20.86\sym{+}&      -87.08\sym{*}&      -81.92\sym{+}&       25.58\sym{+}&      -108.1\sym{*}&      -103.1\sym{*}&       25.67\sym{+}&      -131.0\sym{*}\\
            &     (-1.72)       &      (1.93)       &     (-2.17)       &     (-1.71)       &      (1.67)       &     (-2.18)       &     (-1.98)       &      (1.80)       &     (-2.13)       \\
Graduation - 6-10 years         &      -1.638       &       5.402\sym{*}&      -8.418       &      -0.944       &       10.79\sym{*}&      -13.70       &      -20.09       &       28.73\sym{*}&      -54.58\sym{*}\\
            &     (-0.17)       &      (2.03)       &     (-0.99)       &     (-0.09)       &      (3.25)       &     (-1.31)       &     (-1.44)       &      (3.20)       &     (-2.23)       \\
Graduation -11-15 years          &       12.95       &       8.312       &       3.152       &       13.41       &       10.68\sym{+}&       0.715       &       22.10\sym{*}&       21.02\sym{*}&      -3.771       \\
            &      (0.91)       &      (1.54)       &      (0.62)       &      (1.08)       &      (1.67)       &      (0.14)       &      (2.46)       &      (2.34)       &     (-0.44)       \\
Graduation -31-35 years           &       13.47       &       0.675       &       12.19       &       17.68       &       1.292       &       15.60       &       35.58\sym{+}&       6.513       &       27.81\sym{*}\\
            &      (1.36)       &      (0.61)       &      (1.21)       &      (1.57)       &      (0.76)       &      (1.49)       &      (1.86)       &      (1.24)       &      (1.97)       \\
\hline\hline
\multicolumn{10}{l}{\footnotesize \textit{t} statistics in parentheses}\\
\multicolumn{10}{l}{\footnotesize \sym{+} \(p<0.10\), \sym{*} \(p<0.05\)}\\
\end{tabular}
\caption{Effects by age and graduation cohort.}
\label{dids_time}
\end{table}

\end{landscape}

\begin{table} \centering 
\def\sym#1{\ifmmode^{#1}\else\(^{#1}\)\fi}
\begin{tabular}{l*{1}{ccc}}
\hline\hline

            &        Mean - FMD &          Mean - Other MDs &          p-value\\
\hline
License revoked     &     0.00124&         0.000824&           0.254\\
Terms and conditions applied &      0.0171&          0.0208&          0.0192\\
Suspension of license  &     0.00336&           0.00282&             0.403\\
Resigned from membership    &       0.164&        0.0905&          1.74e-84\\ \hline
N & 13709& 17059& \\
\hline\hline
\end{tabular}
\caption{Summary statistics and t-tests for key quality outcomes by whether the physician is in a primary care or any other specialty.}
\label{cpso_summary}
\end{table}

\begin{table} \centering
\def\sym#1{\ifmmode^{#1}\else\(^{#1}\)\fi}
\begin{tabular}{l*{3}{c}}
\hline\hline
            &\multicolumn{1}{c}{2 year effect}&\multicolumn{1}{c}{4 year effect}&\multicolumn{1}{c}{10 year effect}\\
\hline
Revoked         &     0.00144       &   0.0000814       &    0.000222       \\
            &      (1.23)       &      (0.06)       &      (0.45)       \\
Suspension     &   -0.000503       &    0.000805       &    0.000827       \\
            &     (-0.19)       &      (0.37)       &      (0.40)       \\
Terms and Conditions         &     0.00248       &   -0.000810       &    -0.00317       \\
            &      (0.61)       &     (-0.25)       &     (-1.18)       \\
Resigned from membership         &      0.0246\sym{+}&      0.0143       &     0.00763       \\
            &      (1.93)       &      (1.23)       &      (0.76)       \\
\hline
\(N\)       &        30,703            &       30,703             &     30,703               \\

\hline\hline
\multicolumn{4}{l}{\footnotesize \textit{t} statistics in parentheses}\\
\multicolumn{4}{l}{\footnotesize \sym{+} \(p<0.10\), \sym{*} \(p<0.05\)}\\
\end{tabular}
\caption{Effects of increased primary care residency length on quality measures in Ontario.}
\label{cpso_quality}
\end{table}

\newpage
\FloatBarrier
\appendix
\counterwithin{figure}{section}
\counterwithin{table}{section}

\section{Additional Tables and Figures}

\begin{table}[h!] \scriptsize \centering
\def\sym#1{\ifmmode^{#1}\else\(^{#1}\)\fi}
\begin{tabular}{l|c c c }
\hline\hline
            &\multicolumn{3}{c}{Family Medicine Practitioners}\\
            \hline
            &\multicolumn{1}{c}{Total}&\multicolumn{1}{c}{FMDs}&\multicolumn{1}{c}{GPs}  \\
\hline
\hline
DiD Estimates (2 Years) &      -1.546\sym{+}&       1.485\sym{*}&      -3.108\sym{*} \\
&     (-1.82)       &      (4.00)       &     (-3.74)     \\
\hline
DiD Estimates (4 Years) &      -1.865\sym{+}&       1.934\sym{*}&      -3.946\sym{*}  \\
&     (-1.80)       &      (4.72)       &     (-3.78)\\
\hline
DiD Estimates (10 Years)  &      -0.234       &       3.966\sym{*}&      -4.742\sym{*}  \\
&     (-0.22)       &      (8.09)       &     (-4.40)     \\ \hline
\(N\)       &           8690        &         8690   &       8690    \\
Mean        &       84.34       &       16.34       &       67.35       \\
\hline\hline
\multicolumn{4}{l}{\footnotesize \textit{t} statistics in parentheses}\\
\multicolumn{4}{l}{\footnotesize \sym{+} \(p<0.10\), \sym{*} \(p<0.05\)}\\
\end{tabular}
\caption{DiD Estimates by type of physician. The control group is altered to specialties with lower substitutability and includes surgical and diagnostic specialties. }
\label{table_dids2}
\end{table}

\end{document}